\documentclass[prl,reprint,twocolumn,amsmath,amssymb,nofootinbib]{revtex4-1}

\usepackage{graphicx}
\usepackage{dcolumn}
\usepackage{bm}

\begin{document}

\title{$\mathcal{PT}$-symmetry breaking in resonant tunneling heterostructures}

\author{A.\,A.\,Gorbatsevich}
 \affiliation{P.N. Lebedev Physical Institute of the Russian Academy of Sciences, 119991, Moscow, Russia.}
 \altaffiliation[Also at ]{National Research University of Electronic Technology, 124498, Zelenograd, Moscow, Russia.}
 \email{aagor137@mail.ru}

\author{N.\,M.\,Shubin}%
\affiliation{National Research University of Electronic Technology, 124498, Zelenograd, Moscow, Russia.}

\date{\today}

\begin{abstract}
We present fermionic model based on symmetric resonant tunneling heterostructure, which demonstrates spontaneous symmetry breaking in respect to combined operations of space inversion ($\mathcal{P}$) and time reversal ($\mathcal{T}$). $\mathcal{PT}$-symmetry breaking manifests itself in resonance coalescence (collapse of resonances). We show that resonant energies are determined by eigenvalues of auxiliary pseudo-Hermitian $\mathcal{PT}$-invariant Hamiltonian.
\end{abstract}

\maketitle

Spontaneous symmetry breaking (SSB) is a central concept in different fields of modern physics, especially in particle physics~\cite{bib:SSB} and condensed matter physics~\cite{bib:SSBCM1, bib:SSBCM2, bib:SSBCM3, bib:SSBCM4, bib:SSBCM5}. SSB means that the symmetry of the system changes (lowers) at some value of a system parameter, which itself does not change the symmetry directly. Recently a new class of SSB phenomena was introduced in $\mathcal{PT}$-invariant systems~\cite{bib:PTBend98, bib:PTBend99, bib:PTBend07}. Such systems are invariant with respect to both space inversion ($\mathcal{P}$) and time reversal ($\mathcal{T}$) and are described by $\mathcal{PT}$-invariant pseudo-Hermitian Hamiltonian, which can possess real eigenvalues~\cite{bib:PTBend98}. At some magnitude of Hamiltonian parameter two real eigenvalues coalesce and transform into another two with non-zero imaginary parts of different signs and with equal real parts -- $\mathcal{PT}$-symmetry breaking ($\mathcal{PT}$-SB)~\cite{bib:PTBend98, bib:PTBend99, bib:PTBend07}. Such points in the parameter space are known as exceptional points (EP)~\cite{bib:BookKato, bib:Heiss, bib:Berry}. The eigenstate of the Hamiltonian at EP is nondegenerate (contrary to crossing point). Hamiltonian eigenvalue with positive imaginary part corresponds to nonunitary evolution of wave function, which is forbidden by norm preserving condition. Hence, it  was not clear whether fermionic systems could exist with some relation to pseudo-Hermitian $\mathcal{PT}$-invariant Hamiltonian. Up to now all realistic applications of $\mathcal{PT}$-SB with possible experimental manifestations have been based on the formal equivalence of Schr\"oedinger and wave equations and described electromagnetic phenomena~\cite{bib:PTopt1, bib:PTopt2, bib:Laser, bib:nature, bib:Cho, bib:Ambi}. $\mathcal{T}$-breaking terms in the wave equation correspond to well established gain/loss processes.  Superconducting $\mathcal{PT}$-invariant model was considered in Ref.~\cite{bib:PTsc}. $\mathcal{T}$-breaking terms in this case describe creation/annihilation processes in bosonic Cooper-pair field. 

In this paper we present fermionic model with ($\mathcal{PT}$-SB) based on symmetric resonant tunneling structure (RTS). RTS is a typical example of an open quantum system. SSB phenomenon in open quantum system has been already described  in the early treatments~\cite{bib:CalLeg} based on Caldeira-Legget model where SSB could be attributed to tunneling  suppressed by dissipation. Later in Ref.~\cite{bib:Gor} it was shown that in symmetric RTS without dissipation SSB can occur at the point where two resonances coalesce. This phenomenon was called the collapse of resonances (CR). In this paper we construct auxiliary pseudo-Hermitian $\mathcal{PT}$-invariant Hamiltonian whose eigenvalues exactly correspond to resonance energies and EP describe CR. In condensed matter physics the description and classification of states with broken symmetry are based on group theory, which limits the possible number of different states. In the case of CR in RTS the mirror symmetry is the only symmetry that is broken. However, the number of resonances, which coalesce can vary.

Consider tunneling of electrons through an arbitrary multi-barrier structure. We use Keldysh formalism in tight-binding approximation~\cite{bib:Car1,bib:Car2} that provides a unique description both for coherent and incoherent (dissipation) processes. We begin with the description of coherent transport. Current through the RTS consisting of $N$ coupled wells can be written in the well-known form~\cite{bib:Car1,bib:Kapa}:
\begin{equation}
I=\frac{e}{2\pi}\int{4\Gamma_{L}(\omega)\Gamma_{R}(\omega)|{G}^{r}_{1N}(\omega)|^{2}(f_{L}(\omega)-f_{R}(\omega))d\omega}.
\label{eq0}
\end{equation}
Here $f_{L,R}$ is a Fermi distribution function, ${G}^{r}_{1N}$ is a full retarded Green's function of the system and $\Gamma_{L,R}=\pi|t_{L,R}|^{2}\rho_{L,R}$ is a tunneling rate from outer ($1$-st or $N$-th) well into the left/right contact with density of states $\rho_{L,R}$ via matrix element $t_{L,R}$. According to~(\ref{eq0}) we can define transmission of the structure as~\footnote{subscript $NW$ means that we consider $N$-well RTS.}:
\begin{equation}
T_{NW}=4\Gamma_{L}\Gamma_{R}|{G}^{r}_{1N}|^{2}.
\label{eq01}
\end{equation}
The full propagator ${G}^{r}_{1N}$ accounts for interaction with the continuum in the bulk. Using appropriate contact self-energies~\cite{bib:Car1}:
\begin{equation}
\Sigma_{L,R}=|t_{L,R}|^{2}g_{L,R}^{r}=\delta_{L,R}-i\Gamma_{L,R},
\label{eq02}
\end{equation}
where $g_{L,R}^{r}$ are retarded Green's functions in the contacts, we can follow, for example, Ref.~\cite{bib:Kapa} and write it in the form:
\begin{equation}
{G}_{1N}^r=\frac{G_{1N}^{0r}}{\Delta},
\label{eq3}
\end{equation}
where $G_{1N}^{0r}$ is the retarded Green's function of isolated RTS with no interaction with contacts and
\begin{equation*}
\Delta=(1-\Sigma_{L}G_{11}^{0r})(1-\Sigma_{R}G_{NN}^{0r})-\Sigma_{L}\Sigma_{R}G_{1N}^{0r}G_{N1}^{0r}.\nonumber
\end{equation*}
The retarded Green's function $G_{1N}^{0r}$ of the system of $N$ coupled wells not connected to the bulk can be obtained from the following Dyson equation~\cite{bib:Car1}:
\begin{equation}
G_{ij}^{0r}=\delta_{ij}g_{ii}^r+g_{ii}^r \left(\tau_{i} G_{i+1,j}^{0r}+\tau_{i-1}^{*} G_{i-1,j}^{0r}\right),
\label{eq1}
\end{equation}
where $g_{ii}^r(\omega)=(\omega-\varepsilon_{i}+i0)^{-1}$ are Green's functions of isolated wells with single energy level $\varepsilon_{i}$ and $\tau_{i}$ is a tunneling matrix element between $i$-th and $(i+1)$-th wells.

Using properties of tridiagonal matrix minors decomposition one can show that transmission coefficient~(\ref{eq01}) can be written as a fraction with characteristic polynomial of system's effective Hamiltonian in denominator~\cite{bib:Cel}:
\begin{equation}
T_{NW}=\frac{P^{2}}{\left|\det\left(\omega-\hat H_{eff}\right)\right|^{2}},
\label{eqTNclass}
\end{equation}
where $P^{2}=4\Gamma_{L}\Gamma_{R}|\tau_{1}|^2\cdot...\cdot|\tau_{N-1}|^{2}$. Effective Hamiltonian $\hat H_{eff}$ in~(\ref{eqTNclass}) we write as usual:
\begin{equation}
\hat H_{eff}=\hat H_{0}+\hat H_{L}+\hat H_{R}.
\label{eqHeff}
\end{equation}
Here $\hat H_{0}$ is the Hamiltonian of closed $N$-well system:
\begin{equation}
\hat H_{0}=\begin{pmatrix}
		\varepsilon_{1} & \tau_{1} & \ldots & 0 & 0\\
		\tau_{1}^{*} & \varepsilon_{2} & \ldots & 0 & 0\\
		\vdots & \vdots & \ddots & \vdots & \vdots\\
		0 & 0 & \ldots & \varepsilon_{N-1} & \tau_{N-1}\\
		0 & 0 & \ldots & \tau_{N-1}^{*} &\varepsilon_{N}
		\end{pmatrix},
\label{eqH0}
\end{equation}
and contacts are taken into account by $(\hat H_{L})_{ij}=\Sigma_{L}\delta_{i1}\delta_{j1}$ and $(\hat H_{R})_{ij}=\Sigma_{R}\delta_{iN}\delta_{jN}$. Real parts $\delta_{L,R}$ of self-energies correspond to energy shift and imaginary parts $\Gamma_{L,R}$ describe decay into the bulk's continuum~(\ref{eq02}), close in sense to Feshbach optical potential~\cite{bib:F1,bib:F2}.

Eigenvalues of $\hat H_{eff}$ coincide with poles of scattering matrix~\cite{bib:Res, bib:Dittes}. However, as follows from~\cite{bib:Gor, bib:Res}, unity peaks of transmission do not coincide with these eigenvalues. After some algebra in denominator of~(\ref{eqTNclass}) we can rewrite transmission of arbitrary multi-well system as:
\begin{equation}
T_{NW}=\frac{P^{2}}{|Q|^{2}+P^{2}},
\label{eqTN}
\end{equation}
where
\begin{equation}
Q=\det{\left(\omega-\hat{H}_{aux}\right)}
\label{eqQd}
\end{equation}
is a characteristic polynomial of non-Hermitian auxiliary Hamiltonian describing electron flow from right to left~\footnote{Changing of electron flow direction to from left to right will lead to redefinition $\hat{H}_{aux}=\hat H_{0}+\hat H_{L}+\hat H_{R}^{*}$. Also one can see that electron flow direction has no influence on transmission coefficient, as expected.}:
\begin{equation}
\hat{H}_{aux}=\hat H_{0}+\hat H_{L}^{*}+\hat H_{R}.
\label{eqHeff2}
\end{equation}
Here $H_{L}^{*}$ is a complex conjugate of $H_{L}$ from~(\ref{eqHeff}). Thus, we get, that resonances of transmission are defined by eigenvalues of auxiliary Hamiltonian~(\ref{eqHeff2}), which contrary to $\hat H_{eff}$ from~(\ref{eqHeff}) has different signs of imaginary terms in the first and the last element on the main diagonal. Expressions~(\ref{eqTN}-\ref{eqQd}) represent compact generalization of Breit-Wigner formula to the multilevel case.

From now on we will concentrate our analysis on symmetric $N$-well RTS only ($t_{L}=t_{R}$, $\varepsilon_{i}=\varepsilon_{0}$ and $\tau_{N-i}=\tau_{i}$ for any $i$). Moreover, we assume energy shifts in the outer wells to be $\delta_{L}=\delta_{R}=0$ (physically this means that we consider structures with energy levels in the wells situated in the middle of barriers height). Such an assumption makes the auxiliary Hamiltonian from~(\ref{eqHeff2}) to be $\mathcal{PT}$-symmetric~\footnote{It is invariant with respect to simultaneous time reversal (complex conjugation) and space mirror reflection ($j\in\{1,...,N\}\mapsto N+1-j$).}:
\begin{equation}
\hat{H}_{aux}^{symm}=\hat H_{\mathcal{PT}}=\hat H_{0}+\hat H_{\Gamma},
\label{eqHPT}
\end{equation}
where $(\hat H_{\Gamma})_{ij}=i\Gamma(\delta_{i1}\delta_{j1}-\delta_{iN}\delta_{jN})$ corresponds to the case of electron flow from left to right. Thus, for symmetric RTS, unity peaks of transmission, according to~(\ref{eqTN}), are defined by eigenvalues of non-Hermitian $\mathcal{PT}$-symmetric Hamiltonian~(\ref{eqHPT}). For example, in double-barrier symmetric RTS auxiliary Hamiltonian is a real scalar describing one resonance peak and so $Q_{1W}=\omega-\varepsilon_{0}$. Expression~(\ref{eqTN}) for the transmission in this case turns into simple Breit-Wigner formula. For structures with $N=2,3,4,5$ wells $Q$ polynomials are (we assume $\omega-\varepsilon_{0}\mapsto\omega$):
\begin{equation}
\begin{split}
Q_{2W}=&\omega^{2}-|\tau_{1}|^2+\Gamma^{2},\\
Q_{3W}=&\omega\left(\omega^{2}-2|\tau_{1}|^{2}+\Gamma^{2}\right),\\
Q_{4W}=&\omega^{4}-\omega^{2}\left(2|\tau_{1}|^{2}+|\tau_{2}|^{2}-\Gamma^{2}\right)+\\
&|\tau_{1}|^{4}-|\tau_{2}|^{2}\Gamma^{2},\\
Q_{5W}=&\omega(\omega^{4}-\omega^{2}\left(2|\tau_{1}|^{2}+2|\tau_{2}|^{2}-\Gamma^{2}\right)+\\
&|\tau_{1}|^{4}+2|\tau_{1}|^{2}|\tau_{2}|^{2}-2|\tau_{2}|^{2}\Gamma^{2}).
\label{eqQ}
\end{split}
\end{equation}
At the very moment of CR in $N$-well symmetric RTS transmission coefficient takes significantly non Breit-Wigner form:
\begin{equation}
T_{NW}(\omega)=\frac{\tilde\Gamma^{2N}}{(\omega-\varepsilon_{0})^{2N}+\tilde\Gamma^{2N}}.
\label{eqTNcol}
\end{equation}
where $\tilde\Gamma=C_{N}\Gamma$ and $C_{N}$ is a nonzero constant depending on $N$. The main feature of (\ref{eqTNcol}) is that first $2N-1$ derivatives take zero values at point $\omega=\varepsilon_{0}$ and the first nonzero derivative will be only of the $2N$-th order.

Generally, polynomial  $Q$ has $N-2M$ real and $2M$ complex roots but only real roots correspond to unity values of transmission coefficient. Varying system parameters $\{\tau_{i}\}$ and $\Gamma$ one can make real roots (and so resonances) to coalesce and turn into complex {ones}. In the case of even $N$ all roots of $Q$ can be made complex and so the only remaining transmission peak will be less than $1$ and have asymmetric distribution of electron wavefunction probability. In the case of odd $N$ there will always be at least one real root corresponding to $\omega=\varepsilon_{0}$. After CR this peak will retain its unity magnitude and symmetry of electron wavefunction. According to standard matrix theory~\cite{bib:RealEig} eigenvalues of tridiagonal matrix $\hat H_{0}$ in~(\ref{eqHPT}) are all real and simple. Hence in the case of weak interaction with continuum ($\Gamma\ll\min{(|\tau_{i}|)}$) all resonances are distinct. With increasing $\Gamma$ real eigenvalues can coalesce and transform into complex conjugates, that will result in coalescence of corresponding resonances. Thus, collapse of resonances is obtained at certain ratios of structure parameters when all eigenvalues coalesce together and so CR point is a EP of order $N$. The CR in symmetric structure with even number of wells $N$ in such a treatment corresponds to a $\mathcal{PT}$-SB in auxiliary Hamiltonian $\hat{H}_{aux}$.

From~(\ref{eqQ}) it follows that in double-well structure resonances coalesce at $\Gamma=|\tau_{1}|$, in triple-well case CR occurs at critical ratio of the parameters: $\Gamma=\sqrt{2}|\tau_{1}|$, in 4-well structure critical ratios are: $|\tau_{2}|=|\tau_{1}|\sqrt{\sqrt{2}-1}$, $\Gamma=|\tau_{2}|(\sqrt{2}-1)^{-1}$ and for 5-well RTS: $|\tau_{1}|=|\tau_{2}|\sqrt{1+\sqrt{5}}$, $\Gamma=|\tau_{2}|\sqrt{4+2\sqrt{5}}$. For even $N$ polynomial $Q_{NW}$ is an $N/2$-th order polynomial of $(\omega-\varepsilon_{0})^2$. While for for odd $N$ polynomial $Q_{NW}$ is an $(N-1)/2$-th order polynomial of $(\omega-\varepsilon_{0})^2$ multiplied by $(\omega-\varepsilon_{0})$. This means that the number of independent structure's parameters ($\{\tau_{i}\}$ and $\Gamma$) is enough to totally control the location of the roots of polynomial $Q$. Hence all possible resonance locations can be achieved by varying structure's parameters. Transition from precollapse state of the system (all resonances are separate) into postcollapse state (only one resonance left) can be performed through the point of CR (EP), where all unity resonances coalesce together. To realize this transformation it is enough to change system parameters in some direction in the parameter space. In triple- and four-barrier RTS such a transition can be realized by increasing parameter $\gamma_{2,3W}=\Gamma/\tau_{1}$, (Fig. 1(a,b)). Generally, in more complicated structures required direction is a particular Jordan path in the parameters space and its natural parameter $\gamma_{NW}$ in this space defines how one should tune the system in order to get the CR. The choice of an appropriate path defines interaction between resonances (roots of $Q$) and so one can get all possible variants of coalescence of resonances by choosing different paths. Thus, there are many ways to achieve CR, for example, for $N=2,3,4,5$ possible behavior schemes of transmission peaks positions are shown in Fig. 1.

\begin{figure}
\includegraphics{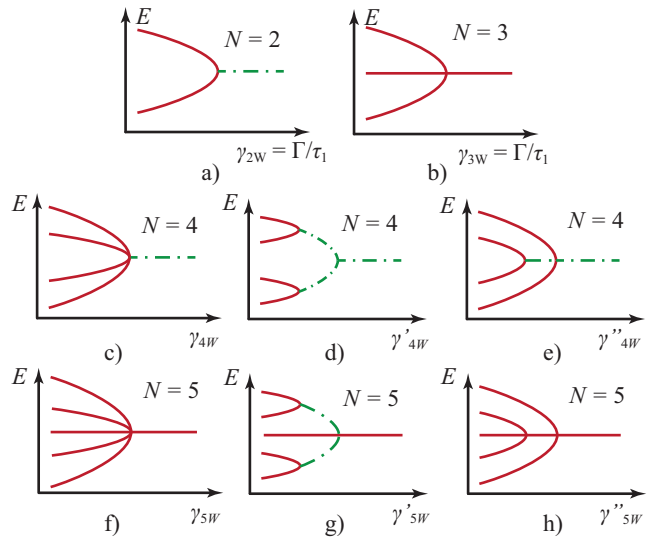}
\caption{\label{fig1}Schematic view of possible types of CR (EP) in RTS with $N=2,3,4,5$ wells. Solid line -- unity valued maximums of transmission coefficient, dot-dashed line -- non-unity maximums.}
\end{figure}

The scenario presented above is based on tight-binding model with nearest neighbors hopping. However, all the qualitative conclusions are quite universal. To demonstrate the key difference between CR in systems with even and odd number of wells $N$, we consider numerical solutions of effective mass Shr\"odinger equation:
\begin{equation}
\left(-(2m^*)^{-1}\hbar^2 \nabla^2+U(x)\right)\psi(x)=E\psi(x),
\label{eqShr}
\end{equation}
where potential $U(x)$ describes multi-well RTS. Figure~2 shows transmission and electron wavefunction distribution in the postcollapse state for structures with $N=2,3,4,5$ wells with barrier height of 0.3 eV, outside barrier thickness of 2 nm and effective mass $m^{*}=0.067 m_{0}$. In the case of even $N$ the postcollapse transparency peak possesses non-unity magnitude and symmetry (of electron wavefunction distribution) is broken, while for odd $N$ the postcollapse peak transparency is perfect and symmetry is conserved. Mechanism of symmetry breaking at the CR point is quite simple. Electron wave functions corresponding to adjacent levels near CR point differ in symmetry. At the CR point energy levels coalesce into a single non-degenerate one, which wave function is the combination of symmetric and antisymmetric functions and hence is non-symmetric.

\begin{figure}
\includegraphics{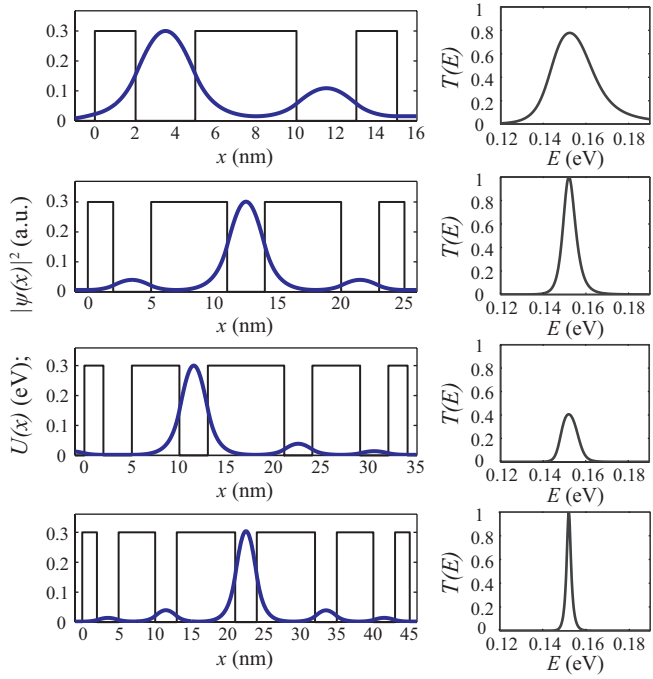}
\caption{\label{fig2}Numerically calculated electron distributions and transmission coefficients for symmetric RTS with $N=2,3,4,5$ wells in the postcollapse state ($\gamma_{(N+1)B}>\gamma_{(N+1)B}^{crit.}$).}
\end{figure}

To summarize, we have shown that $\mathcal{PT}$-symmetry breaking (SB) exists in symmetric quantum multi-well structure with even number of wells. $\mathcal{PT}$-SB manifests itself as the coalescence of resonances. We have introduced auxiliary pseudo-Hermitian $\mathcal{PT}$-symmetric Hamiltonian $H_{\mathcal{PT}}$ whose eigenvalues determine exactly the positions of resonance maxima. $H_{\mathcal{PT}}$ can be straightforwardly deduced from effective optical-potential-like Hamiltonian describing decaying states of open quantum system. $\mathcal{PT}$-SB corresponds to exceptional points of auxiliary Hamiltonian $H_{\mathcal{PT}}$. Thus, $\mathcal{PT}$-SB in fermionic system has been described. Exceptional points in structures with odd number of quantum wells are not accompanied by $\mathcal{PT}$-SB. Optical counterpart of $\mathcal{PT}$-SB described in the present paper could be observed in coupled optical wave-guides.

\begin{acknowledgments}
AAG would like to acknowledge the Program of Fundamental Research of the Presidium of the Russian Academy of Science for partial support. 
\end{acknowledgments}

\bibliography{refer}

\end{document}